\nopagenumbers
\def \half {{\textstyle {1\over 2}}}
\def \a {{\alpha}}
\def \Tr {{\rm Tr}}
\def \bx {{\bar \xi}}
\def \bZ {{\bar Z}}
\def \bone {{\bar 1}}
\def \btwo {{\bar 2}}
\magnification =\magstep 1
\overfullrule=0pt
\hsize = 6.5truein
\vsize = 8.5truein
\null
\vskip-1truecm
\rightline{IC/98/17}
\rightline{CCNY-HEP 98/1}
\vskip1truecm
\centerline{
United Nations Educational Scientific and Cultural Organization}
\centerline{and}
\centerline{International Atomic Energy Agency} \medskip
\centerline{
INTERNATIONAL CENTRE FOR THEORETICAL PHYSICS}
\vskip1.5truecm
\centerline{\bf On  brane solutions in M(atrix) theory}
\vskip 1.5cm
\centerline{
V.P. Nair$^1$  and S. Randjbar-Daemi$^2$}
\vskip .2 in
\centerline{$^1$ City College of the CUNY, New York, NY 10031}
\vskip .05in
\centerline{$^2$ Abdus Salam International Centre for Theoretical Physics}
\centerline{Trieste, Italy}
\vskip 1.2cm
\centerline{ABSTRACT}
\baselineskip=18pt
\vskip .1in
In this paper we consider brane solutions of the form $G/H$ in M(atrix)
theory,
showing  the emergence of world volume
coordinates for the cases where $G=SU(n)$.
We examine a particular solution with a world volume geometry of the form
${\bf CP}^2\times S^1$ in some detail and show how a smooth
manifold structure emerges in the large
$N$ limit. In this limit the solution becomes
static; it is not supersymmetric but is part of a supersymmetric set of
configurations. Supersymmetry in small locally flat regions can be obtained,
but this is not globally defined.
A general group theoretic analysis
of the previously known spherical brane solutions is also given.
\vskip 1.2cm
\centerline{MIRAMARE -- TRIESTE}
\centerline{February 1998}
\vfill\eject
\footline{\hss\tenrm\folio\hss}
\pageno=2
\noindent{\bf 1. Introduction}
\vskip .1in
The matrix theory proposed by BFSS [1, 2] as a version of M-theory has been
rather intensively investigated over the last year or so.
It is by now clear that it does
capture many of the expected features of M-theory such as the
11-dimensional supergravity regime and the existence of
extended objects of appropriate dimensions, although there are still
some issues to be resolved regarding recently found discrepancies [3,4].
Smooth extended objects pertain to the large $N$-limit as well as
nonperturbative regimes of the theory and as such, the study of
these objects, particularly of curved brane solutions, is of interest.
The emergence of the two-brane or the standard membrane was analyzed
many years ago in the paper of de Wit, Hoppe and Nicolai [5].
More recently, spherically symmetric membranes have been obtained [6,7].
As regards the five-brane, which is the other extended object of
interest, there has been no satisfactory construction or understanding
of the transverse brane where all five spatial dimensions are a subset
of the nine manifest dimensions of the matrix theory. The longitudinal
five-branes, called $L5$-branes, which have four manifest dimensions
and one along the compactified direction (either the 11-th dimension
or the lightlike circle) have been obtained. These include flat
branes [8,9] and stacks of $S^4\times S^1$-branes [10]. In this paper
we analyze the construction of brane configurations. We obtain a group
theoretical rephrasing of the known solutions and some new solutions
such as an $L5$-brane of ${\bf CP}^2\times S^1$-geometry. For this latter
case it is possible to have a single smooth static brane configuration,
rather than stacks of them, in the large $N$-limit. Static {\it solutions},
as opposed to merely static configurations, are possible if
$N\rightarrow \infty$ with $R$ fixed, where $R$ is the radius of the
lightlike circle or the 11-th dimension.

In our analysis, the condition for finite
potential energy in the large $N$-limit can be easily obtained
in a group theoretic way. The static solution we obtain
has finite energy as $N\rightarrow \infty$.
In general, the geometry of curved brane
solutions must clearly be constrained by the requirement that
they be solutions
in M-theory. The fact that one can obtain finite static energy
only for a small class of configurations is presumably related to a
M(atrix) theory version of this condition.

The plan of this paper is as follows: In section two we explain our ansatz
and its group theoretical structure. In this section we also show
how the known spherical branes have  a simple description in the framework of
our analysis. Section three is devoted to the discussion of
one particular solution, namely, a five-brane with the world volume
geometry of ${\bf CP}^2\times S^1$.
In this section we also indicate the way to introduce local coordinates
in the world volume  of the brane and discuss the emergence of the smooth
manifold
structure in the large $N$ limit. The immersion of the ${\bf CP}^2$
in ${\bf R}^9$ is also discussed. In section four we discuss
the response of the  ${\bf CP}^2\times S^1$ solution to supersymmetry
transformations. Our
solution is not
supersymmetric, but is part of a set of degenerate configurations which are
related
by supersymmetry. We argue
that it also admits some unbroken $N=4$ supersymmetry in small locally flat
neighborhoods; this notion is however not globally defined and hence the
implications
of this symmetry are unclear.
In section five we discuss some other solutions and give a different
descriptions of
the spherical membrane. The paper concludes
with a
brief discussion.

\vskip.1in
\noindent{\bf 2. The ansatz}
\vskip .1in
The matrix theory Lagrangian can be written as
$$
{\cal L}= {{\Tr}} \left[ {{\dot X}_I^2\over 2R} +{R\over 4}[X_I,X_J]^2
+\theta^T{\dot \theta}+i R \theta^T \Gamma_I [X_I,\theta]\right]
\eqno(1)
$$
where $I,J=1,...,9$ and $\theta$ is a $16$-component spinor
of $O(9)$ and $\Gamma_I$ are the appropriate gamma matrices.
$X_I$ are hermitian $(N\times N)$-matrices; they are elements
of the Lie algebra of $U(N)$ in the fundamental representation.
The theory is defined by this Lagrangian supplemented by the
Gauss law constraint
$$
[X_I,{\dot X}_I] -[\theta, \theta^T] \approx 0 \eqno(2)
$$
In the following we shall be concerned with bosonic
solutions and the $\theta$'s will be set to zero.
The relevant equations of motion are thus
$$
{1\over R}{\ddot X}_I ~-~ R [X_J,[X_I,X_J]] =0
\eqno(3)
$$

We shall look for solutions which carry some amount of symmetry.
In this case, simple ans\"atze can be formulated in terms of a
group coset space $G/H$ where $H\subset G \subset U(N)$.
The ans\"atze we consider will have spacetime symmetries,
a spacetime transformation being compensated by an
$H$-transformation. In the large $N$-limit, the matrices $X_I$
will go over to continuous brane-like solutions with the geometry
of $G/H$.

Let $t^A,~A=1,...,dimG$ denote a basis of the Lie algebra of
$G$. We split this set of generators into two groups, $t^\alpha,
~ \alpha =1,...,dimH$ which form the Lie algebra of $H$ and
$t_i,~i=1,..., (dimG-dimH)$ which form the complementary
set. Our ansatz will be to take $X_I$'s to lie along the
entire algebra of ${\underline G}$ or to be  linear combinations
of the $t_i$'s. In the latter case, in order to satisfy the equation of motion
(3), we shall then need the double commutator $[X_J,[X_I,X_J]]$
to be combinations of the $t_i$'s themselves. This is guaranteed
if $[t_i,t_j]\subset {\underline H}$, since the $t_i$'s themselves
transform as representations of $H$.
In this  case we have
$$\eqalignno{
[t^\alpha ,t^\beta ]= if^{\alpha \beta \gamma } t^\gamma,~~~~~~~~&~~~~~~~
[t^\alpha ,t_i] = i f^{\alpha }_{ij} t_j&(4a)\cr
[t_i,t_j] &= ic_{ij}^\alpha t^\alpha &(4b)\cr}
$$
and $G/H$ is a symmetric space. If $H=1$
the $t_i$'s will belong to the full algebra ${\underline G}$.
In this case, $c_{ij}^\a$ of Eq.(4b)
will be the structure constants of ${\underline G}$.
In such a case, eventhough the ansatz involves the full algebra
${\underline G}$, it can satisfy further algebraic constraints which have only
a smaller invariance group $H$. The solution will then again reduce to
the $G/H$-type.  We shall see an example of this in the next section.
These are the cases we analyze.

Since $X_I$ are elements of the Lie algebra of $U(N)$,
having chosen a $G$ and an $H$, we must consider the
embedding of $G$ in $U(N)$.
This is done as follows. We consider a value of $N$ which
corresponds to the dimension of a unitary irreducible
representation (UIR) of $G$. The embedding is then specified
by identifying the fundamental $N$-dimensional representation
of $U(N)$ with the $N$-dimensional UIR of $G$. Eventually we would
want to consider the limit $N\rightarrow \infty$ as well.
Thus we need to consider an infinite sequence of UIR's of
$G$. Different choices of such sequences are possible,
presumably corresponding to different ways of defining the
$N\rightarrow \infty$ limit. A simple convenient choice is to
take the symmetric rank $s$-tensors of $G$, of dimension,
say, $d(s)$. Thus we choose $N=d(s)$, defining the large
$N$-limit by $s\rightarrow \infty$.

The ansatz we take is of the form
$$
X_i= r(t) {t_i\over N^a} \eqno(5)
$$
for a subset $i=1,...,p$ of the nine $X$'s. $t_i$ are generators
in ${\underline G}- {\underline H}$, in the symmetric rank $s$-tensor
representation of $G$. The eigenvalues fo any of the $t_i$'s in the
$s$-tensor representation will range from $cs$ to $-cs$, where
$c$ is a constant. The eigenvalues thus become dense with a
finite range of the variation of the $X_i$'s  as $s\rightarrow \infty$
if $N^a \sim s$. In this case, as $s\rightarrow \infty$, the $X_i$'s
will tend to a smooth brane-like configuration. Our choice of the
index $a$ will be fixed by this requirement, viz., $N^a\sim s$.
Notice that this ansatz is consistent with the Gauss law (2).

The ansatz (5) has a symmetry of the form
$$
R_{ij} ~U X_j U^{-1} = X_i \eqno(6)
$$
where $R_{ij}$ is a spatial rotation of the $X_i$'s and $U$ is
an $H$-transformation for the $G/H$ case or more generally it can be in
$U(N)$. $R_{ij}$ is determined by
the choice of the $X_i$'s involved in (5). Further, the ansatz (5) is to be
interpreted as being given in a specific gauge. A $U(N)$-transformation,
common to all the $X_i$'s, is a gauge transformation and does not
bring in new degrees of freedom. We may alternatively say that the meaning of
Eq.(6) is that $X_j$ is invariant under rotations $R_{ij}$ upto a gauge
transformation.

For the ansatz (5), the Lagrangian simplifies to
$$
{\cal L}= A_s \left[ {{\dot r}^2 \over 2RN^{2a}} ~-~
{{c_{ij\alpha}c_{ij\alpha}\over 4 N^{4a}}}Rr^4 \right]
\eqno(7)
$$
where $A_s$ is defined \footnote
{*}{We normalize the generators of $G$ such that in the fundamental
representation
 ${\Tr} ( t_At_B) = {1\over 2}\delta_{AB}$ . }
 by ${\Tr} ( t_At_B) =A_s \delta_{AB}$ and the
matrices and trace are in the $s$-tensor of $G$.
{}From its definition,
$A_s = d(s) c_2(s)/dimG$, where $c_2(s)$ is the quadratic Casimir
of the $s$-tensor representation which goes like $s^2$ for large $s$.
Thus, with $N^a\sim s$, the
kinetic term in (7) will always go like
$N/R~$ for large $s$.

We now turn to some specific cases. Consider first $G=SU(2)$.
In this case, $d(s)=s+1,~A_s= s(s+1)(s+2)/12$. The smooth brane
limit thus requires $a=1$ or $N^a \sim N \sim s$.  The kinetic
energy term in (7) goes like $N/R$, while the potential term
goes like $R/N$. Thus both these terms would have a finite
limit if we take $N\rightarrow \infty,~R\rightarrow \infty$,
keeping $(N/R)$ fixed. In fact, this particular property holds
only for $G=SU(2)$ or products thereof, such as $G=O(4)$. For this
reason some of the branes which are realized as
cosets of products of the $SU(2)$ group can perhaps be regarded
as being transverse as their energy will not depend on $R$
in the large $R$ limit.
We can use this case to obtain a slightly
different description of the spherical membrane of [6,7] as well as
 a "squashed" $S^2$ or ${\bf CP}^1$.
The round $CP^1$ corresponds to the case
where the three generators of $SU(2)$ lie along three of the
nine $X_I's$.

As another example, consider $G=O(6)\sim SU(4)$. In this case,
$d(s)= (s+1)(s+2)(s+3)/6, A_s= (1/240)(s+4)!/(s-1)!$ and we need
$a= {1\over 3}$. The kinetic energy term goes like $N/R$
while the potential energy goes like $s R$. This corresponds to the
longitudinal five-brane with
$S^4$-geometry discussed in [10].
$SU(4)$ has an $O(5)$ subgroup under which the $15$-dimensional
adjoint representation of $SU(4)$ splits into the adjoint
of $O(5)$ and the $5$-dimensional vector representation.
The coset generators, corresponding to the $5$-dimensional
representation, may be represented by the $(4\times 4)$-gamma
matrices, $\gamma_i, ~i=1,...,5$.
The ansatz (5) thus takes the form
$X_i = r \gamma_i /N^{1\over 3}$. It is shown
in [9] that the sum of the squares of the $X_i$'s is
proportional to the identity, thus giving effectively a
four-dimensional brane. This is interpreted as $s$ copies of
a longitudinal five-brane, one of the directions being along the
compactified $11$-th or lightcone coordinate, of extent $R$,
an interpretation consistent with
the potential energy $\sim s R$.

An interesting special case which we shall analyze in some detail
corresponds to $G= SU(3)$. In this case, $d(s) = (s+1)(s+2)/2 ~\sim s^2$
and $A_s= (1/48) (s+3)!/(s-1)!$. We choose $a=\half$. The kinetic
energy again goes like $N/R$ for large $N$ while the potential
energy goes like $R$. The potential energy is independent
of $s$ and thus, for $r$ independent of $t$, we have a single
static smooth five-brane wrapped around the compactified dimension
in the $s\rightarrow \infty$ limit. In other words, the tension
defined by the static energy per unit volume remains finite as
$s\rightarrow \infty$.
An appropriate choice of $H$ in this case is $H= U(2)\sim SU(2)\times
U(1)$. The emergent world volume geometry is then
${\bf CP}^2\times S^1$, whose immersion in ${\bf R}^9$ can be complicated
and will depend on the
details of the ansatz. As we shall see in the next section
the coset embedding will produce a singular surface in a 9-dimensional
Euclidean space, while when the eight generators of
$SU(3)$ are set parallel to eight of nine $X_I$'s we shall
obtain the standard $CP^2$ embedded in an $S^7$ contained in $R^9$.

Notice also that since
the effective mass for the degree of freedom corresponding to $r$ goes like
$N/R$,
oscillations
in $r$ are suppressed as $N\rightarrow \infty$ with $R$ fixed; in this limit
this configuration becomes a static {\it solution}. We can fix $r$ to any value
and time-evolution does not change this.
\vskip .1in
\noindent{\bf 3. World volume coordinates and analysis of
${\bf CP}^2\times S^1$-brane}
\vskip .1in
We shall now analyze the ${\bf CP}^2\times S^1$ solution in some detail.
The key to interpreting this as a five-brane is the emergence of the world
volume coordinates in the large $N$-limit. In this limit, the matrices
$X_i$ become truly infinite dimensional and one has something like a
semiclassical limit, the configurations being smooth and matrix commutators
becoming Poisson brackets. The emergence of the world volume coordinates
can be seen in a
suitable parametrization of the generators $t_i$. (Notice that
the energy of a static
configuration, given by the potential energy in Eq.(7),
goes like $Rr^4$, which is naively suggestive of a five-dimensional
extended object. However, this behaviour of the potential energy is identical
for all configurations, i.e., for any kind of extended object,
and one cannot identify the
dimensionality, much less any geometry, from this consideration.)

Since $SU(3)/U(2)$ is ${\bf CP}^2$, we start with a parametrization of the
$SU(3)$ generators in terms of homogeneous coordinates on ${\bf CP}^2$.
Introduce three complex numbers $\pi_\a ,~\a =1,2,3$. The identification
$\pi_\a \sim \lambda \pi_\a,~ \lambda\in {\bf C}-\{0\}$ gives ${\bf CP}^2$.
As homogeneous coordinates, valid for a coordinate patch around $\pi_3 \neq
0$,
we can use $\xi_i = \pi_i /\pi_3,~i=1,2$.
The generators belonging to ${\underline {SU(3)}}-{\underline {U(2)}}$ are
given by
$$
t_i= \pi_3 {\partial \over \partial \pi_i},~~~~~~~~~~~~~~~t_i^\dagger
= \pi_i {\partial \over \partial \pi_3} \eqno(8)
$$
We are interested in the $s$-tensor representation which has the form
$\pi_{\a_1}...\pi_{\a_s}~= \pi_3^s ~f(\xi)$. Simplifying the
above expressions, we find for these states,
$$\eqalign{
t_i=& {\partial \over \partial \xi_i},~~~~~~~~~~~~~~~~t_i^\dagger = \xi_i
\left(
s- \xi_k {\partial \over \partial \xi_k}\right)\cr
h_{ij} \equiv &[t_i,t_j^\dagger ]= \left( s- \xi_k{\partial \over \partial
\xi_k}\right)
\delta_{ij} -\xi_j {\partial \over \partial \xi_i}\cr}
\eqno(9)
$$
The states of the $s$-tensor representation are of the form
$f=~1,~\xi_{i_1},...,
(\xi_{i_1}...\xi_{i_s})$. The scalar product for these states is given by
$$
\langle f \vert g\rangle = {(s+1)(s+2)\over 2}~\int {2~d^4\xi \over {\pi^2
(1+{\bar \xi}\cdot \xi )^3}}~ {1\over (1+{\bar \xi}\cdot  \xi)^s}~{\bar f}g
\eqno(10)
$$
where $\bx \cdot \xi =\bx_k \xi_k$.
$t_i,~t_i^\dagger$ are adjoints with this inner product. Notice also that
finiteness of norm restricts the powers of $\xi_i$ to be
less than or equal to
$s$, giving
$N= d(s)= \half (s+1)(s+2)$ states in all. Also, in matrix elements of
these operators
we can use
$$\eqalign{
t_i &= (s+3){ \bx_i\over {(1+\bx \cdot \xi )}},~~~~~~~~~~~~~~~~t_i^\dagger =
(s+3) {\xi_i \over {(1+\bx \cdot \xi )}}\cr
h_{ij} &= (s+3){(\delta_{ij} -\bx_i \xi_j )\over { (1+\bx \cdot \xi )}}\cr}
\eqno(11)
$$
 To see the emergence of the continuous world volume coordinates
from the matrix model, let us look, for example,
 at the  states near the state $f= 1 \sim \pi_3^s$. On such states we have
$[{t_i\over \sqrt{N}},
{t_j^\dagger \over \sqrt{N}}]\approx
{s\over N} \delta_{ij}$; we can thus use semiclassical simplifications as
$s\rightarrow \infty ,~s/N \rightarrow 0$.
\footnote
{*}{This is similar to the large spin approximation in
Holstein-Primakoff formulation
of magnetic systems in condensed matter physics. }

In this case,  any operator involving products
of $SU(3)$ generators, $t_i, t_i^\dagger, h_{ij}$, etc., may be replaced by
the $c$-number versions as in Eq.(11). The trace of an operator ${\cal O}$
is given by
$$
{\Tr} {\cal O} = {(s+1)(s+2)\over 2}~\int {2~d^4\xi \over {\pi^2
(1+{\bar \xi} \cdot \xi )^3}}~{\cal O}(\bx ,\xi )\eqno(12)
$$
Again, as $s\rightarrow \infty$,
the matrix commutator of any two operators $F,G$ goes over to a Poisson
bracket of the corresponding functions given by
$$\eqalign{
[F,G] \rightarrow \{ F,G \} &= (\Omega^{-1})^{{\bar a} a}(\partial_{\bar a} F
 \partial_a G -\partial_a F \partial_{\bar a}G )\cr
(\Omega^{-1})^{{\bar a} a}&= {(1+\bx \cdot \xi )\over {s+3}}~(\delta^{{\bar
a} a} +
\bx^{\bar a} \xi^a)\cr}
\eqno(13)
$$
$(\Omega^{-1})^{{\bar a} a}$ is the inverse of the K\"ahler form of ${\bf
CP}^2$.

The ansatz for the five-brane may now be stated as follows.
The simplest case to consider is the following.
We define the complex combinations $Z_i = {X_i +i X_{i+2}\over
\sqrt{2}},~i=1,2$.
The symmetric ansatz is then given by
$$\eqalign{
Z_i &= r(t) {t_i\over \sqrt{N}},~~~~~~~~~~~~~~i=1,2 \cr
X_i&=0, ~~~~~~~~~~~~~~~~~~~i=5,...,9\cr}\eqno(14)
$$
In the large $s$-limit, $Z_i\approx r  {{{s}\over{\sqrt {N}}}}
{{\bx_i} \over {(1+\bx \cdot \xi )}}$,
realizing a continuous map from ${\bf CP}^2$ to the space ${\bf R}^9$.
This map is not one-to-one; the region $\bx \cdot \xi <1$ and the region
$\bx \cdot \xi >1$ are mapped into the same spatial region
$\vert Z \vert < r\sqrt{2}$, corresponding to a somewhat squashed ${\bf CP}^2$.

The standard ${\bf CP}^2$ is obtained by
considering an ${\bf R}^8$-subspace
of ${\bf R}^9$, whose coordinates can be identified with the $SU(3)$ generators
as in Eq.(5), i.e., $X_A= rt_A/{\sqrt{N}},~A=1,2,...,8, ~X_9=0$.
Specifically in
 the parametrization of the $SU(3)$ generators in terms of
the $\xi$'s as in Eq.(11),
this ansatz is given by ( recall that in the large $N$ limit
$s\approx \sqrt{2N}$)
$$\eqalign{
X_i&= {r\over\sqrt{2}}~ {{\bar\xi \sigma^i \xi}\over{1 + \bar\xi\xi}}\cr
X_4+iX_5&= {r\over\sqrt{2}}  ~{{ 2 \xi_1}\over{1 + \bar\xi\xi}}\cr
X_6 +iX_7&= {r\over \sqrt{2}} ~ {{ 2\xi_2 }\over{1 + \bar\xi\xi}}\cr
X_8&={{r}\over {\sqrt{6}}}~ {{2 - \xi\bar\xi}\over{1 +
\bar\xi\xi}}\cr
X_9 &=0\cr}\eqno(15a)
$$
where  $\sigma^i$, $i=1,2,3$ are Pauli matrices.
Notice that the singularity of the squashed ${\bf CP}^2$ is removed by this
ansatz since the regions $\bx \cdot \xi <1$ and $\bx \cdot \xi >1$
are mapped to different regions  of the nine-dimensional
space. In fact it
is easy
to see by a direct inspection that the map is actually one-to-one.
Furthermore it is
not hard to verify that $\Sigma_{A=1}^{8} X_A X_A = {2 r^2\over {3}}$. Thus
our surface is embedded in an $S^7$.
In fact we can use  the above relations
and express
$X^1, X^2$ and $X^3$ in terms of $ X^4,.. .,  X^8$ as
$$
X^a= {3\over\sqrt{2}}~{{ \bar\zeta \sigma^a\zeta }\over{r + \sqrt{6} X_8}}
\eqno(15b)
$$
where $\zeta$ is a two-component vector defined by $ \zeta^{1}= {1\over
\sqrt{2}}( X_4+iX_5 )$
and $ \zeta^{2}= {1\over \sqrt{2}}( X_6+iX_7 )$.

To see how the smooth ${\bf CP}^2$ emerges from this construction,
define the
$(3\times 3)$
hermitian traceless matrix $H= \Sigma_{A=1}^8 X_A\lambda_A$,
where $\lambda_A$ are the eight hermitian generators
of $SU(3)$ in the fundamental representation and the $X's$ are
parametrized as above.
By a somewhat lengthy but straightforward
calculation we can show that $H$
satisfies the following
constraint
$$
H^2= {r\over 3\sqrt{2}} H + {r^2\over 9}\eqno(16)
$$
The manifest $SU(3)$ covariance of this equation indicates that our
four-manifold is indeed $SU(3)$ invariant.
In fact $SU(3)$ acts transitively on the surface defined (16), because
any $H$ satisfying the constraint (16) can be obtained from a
diagonal one
given by
$H_{diag.}= {(r/ 3\sqrt{2})}$  diag $( -1, -1,  2)$
by the action of an $SU(3)$-transformation. Furthermore $H_{diag}$ is
invariant under a $SU(2)\times U(1)$ subgroup
of $SU(3)$. It can thus depend only on four real parameters. These
properties are
sufficient to identify the manifold covered by the two complex
$\xi$'s as the
standard smooth ${\bf CP}^2$ embedded in a one-to-one way
in an $S^7$ subspace of
${\bf R}^9$.

The coset structure is clearer directly in terms of the ansatz
for (14), which is why we started with this squashed ${\bf CP}^2$.
The smooth configuration (15) may be regarded as a relaxation of
(14) along some of the ${\bf R}^9$-directions.

A comment regarding the large $s$ limit is in order.
The identification of the generators $t_i$
with the functions of
$\bx, \xi$
as in Eq.(11) can be done for finite $N$ for matrix elements of single powers
of $t_i$.
Furthermore the isomorphism between the $SU(3)$ Lie algebra
 and the algebra defined
by the Poisson brackets is valid for any $s$. However, this isomorphism
extends to the
envelopping algebras, or general functions of the generators,
only in the large $s$ limit.
Therefore it is only for large $s$ that the Lie bracket operation
on any function of the generators of $SU(3)$ will go over
to the Poisson brackets of the corresponding classical functions.
Also, the states are limited to the $s$-th power of the $\xi$'s
by the condition of the finiteness of the norm
and one can get arbitrary world volume deformations of the five-brane
(which involve arbitrarily high powers of $\xi$'s)
only for $s \rightarrow \infty$.
Thus it is only in this limit that
we can expect a continuous five-brane which is an immersion of
${\bf CP}^2$ in spacetime.

The Lagrangian for ans\"atze (14,15) becomes
$$\eqalign{
{\cal L}&= {(s+3)!\over {(s-1)!}}\left[ {{\dot r}^2\over 12 NR} ~-~ {Rr^4
\over 8N^2}
\right]\cr
&\approx \left[ {N\over R} {{\dot r}^2\over 3} ~-~ {Rr^4 \over 2}\right]
(1+{\cal O}({1\over s} ))\cr}\eqno(17)
$$
In terms of the world volume coordinates, we can also write
$$\eqalignno{
{\cal L} &\approx \int {2~d^4\xi \over {\pi^2 (1+\bx \cdot \xi )^3}}
\left[ {s^4\over 2 NR} {\dot r}^2 {\bx \cdot \xi \over {(1+\bx \cdot \xi )^2}}
~-~ Rr^4 {s^4\over 4N^2} {2-2\bx \cdot \xi +(\bx \cdot \xi )^2 \over
{(1+\bx \cdot
\xi )^2}}\right]&(18a)\cr
&\approx \int {2~d^4\xi \over {\pi^2 (1+\bx \cdot \xi )^3}} \left[ {N\over
R}{{\dot r}^2
\over 3}
- {Rr^4\over 2}\right] &(18b)\cr}
$$
Expression (18a) applies to ansatz (14), expression (18b) to ansatz
(15). The energy densities are uniformly distributed over the world volume
for (15), but not for (14).

The equation of motion for $r$ becomes
$$
{N\over R} {\ddot f} +6 f^3 =0 \eqno(19)
$$
where $r= f/{\sqrt{R}}$. The effective mass for the degree of freedom
corresponding to $r$ is ${\textstyle {2\over 3}} (N/R)$. Thus in the
limit $N\rightarrow \infty$ with $R$ fixed, any solution with finite
energy would have to have
a constant $r$ or $f$. In this limit, we thus get a five-brane which is
a {\it static  solution} of
the matrix theory. Alternatively, if we consider $N,R\rightarrow \infty$ with
$(N/R)$ fixed, $f$ can have a finite value.
However, in this limit, the physical dimension of the
brane as given by $r$ would vanish.

Explicit solutions to Eq.(19)
may be written in terms of the  sine-lemniscate function as
$$
f = A \sin {\rm lemn} \left( \sqrt{3R\over N} ~A (t-t_0)\right)
\eqno(20)
$$

The supersymmetry algebra calculation identifies the brane charge as [7]
$$\eqalign{
Q &= {R\over 4!} {\Tr} (X_i X_j X_k X_l) \epsilon^{ijkl}\cr
&= {R\over 12} {\Tr} \left( [Z_1,\bZ_{\bone}][Z_2,\bZ_\btwo ]
-[Z_1,\bZ_\btwo ] [Z_2,\bZ_\bone ]
- [Z_1,Z_2] [\bZ_{\bone} , \bZ_{\btwo} ] \right)\cr}\eqno(21)
$$
For the rest of this section, we shall consider the squashed ${\bf CP}^2$ of
Eq.(14); we have written out $Q$ for this case in Eq.(21).
In the large $s$-limit we find
$$
Q \approx {Rr^4 \over 24} {s^4\over N^2} \int {2~d^4\xi \over {\pi^2
(1+{\bar \xi}\cdot \xi )^3}}~{{1-\bx \cdot\xi}\over (1+\bx \cdot\xi
)^2}\eqno(22)
$$
We see that there is nonuniformly distributed charge density on the world
volume
of the brane. Of course, if the integral
is evaluated, the result will be zero, in consistency with the
evaluation of the trace in Eq.(21). This trace will also vanish for the
$L5$-brane discussed in [10].
This is a reflection of the fact that there is no good
definition of $Q$ for finite dimensional matrices;
it is only as $N\rightarrow \infty$ that a charge can be expected.
In reference [10], a nonzero charge was obtained essentially
by restricting the trace to a subset of states. A similar
definition of
$Q$ will give a nonzero result for our solution as well. For example,
if in the evaluation of the trace we consider only the contributions
coming from the states very near $f= 1 \sim \pi_3^s$,
which corresponds to
the approximation $(1-\bx \cdot\xi )/ (1+\bx \cdot \xi )^2 \approx 1$
in the integrand of (22), we obtain $Q\approx
{{Rr^4}\over{6}}$.

A similar result will hold for the regular ${\bf CP}^2$-solution
of Eq.(15). In this case, we need to choose a subset of four out of the
eight $X_A$'s for use in the formula (21). For $\xi$'s near
zero, the local coordinates correspond to $X_4,X_5,X_6,X_7$.
If this set is used, the result is the same as in Eq.(22), except
for an additional factor of ${1\over 4}$ since
$\zeta_i \sim (r/2 \sqrt{N} )t_i$.

As regards spacetime properties of the solution, the energy we have
evaluated is
the lightcone component $T^{+-}$. Other components can be evaluated,
following the
general formula of [11], and they will act as a source for gravitons, again
along
the general lines of [11]. The current ${\cal T}^{IJK}$ which is the
source for the antisymmetric tensor field of eleven-dimensional
supergravity is another
quantity of interest. This vanishes for the spherically symmetric
configurations
considered in [10], since there are no invariant
$O(9)$-tensors of the appropriate rank and symmetry. However, for the ${\bf
CP}^2$
geometry, there is the K\"ahler form and the possibility that
${\cal T}^{-ij}$ can be proportional to the K\"ahler form has to be
checked explicitly.
The kinetic terms of ${\cal T}^{-ij}$  which depend on ${\dot r}$
are easily seen to vanish for the solution (15), essentially because of
the symmetric nature of the ansatz. For solution (14), we find, by
direct evaluation,
$$\eqalign{
{\cal T}^{-1 \bone} &={\cal T}^{-2\btwo} = -{i\over 60}{s^5\over N^2}
{\dot r}^2 r^2~~+...\cr
{\cal T}^{-1\btwo}&= {\cal T}^{-2\bone}=0\cr}
\eqno(23)
$$
Naively, this diverges as $s\rightarrow \infty$. However, as we have
noticed before,
in this limit, the solution becomes static, ${\dot r}=0$ and hence this
vanishes.
This holds for other components of ${\cal T}^{IJK}$ as well.
The nonkinetic terms in ${\cal T}^{IJK}$ are of the form
$R^2 r^6 ({s^5\over N^3})$ and also vanish as $s\rightarrow \infty$.
Thus the source for the antisymmetric tensor field is zero in the
$s ~({\rm or}~ N)\rightarrow \infty$ limit.

In the alternative limit with $N/R$ fixed, if we write
$r=f/\sqrt{R}$, $f$ has a finite limit and
${\cal T}^{-ij}$ of Eq.(23)
vanishes as $R\rightarrow \infty$,
eliminating possible radiation of the antisymmetric tensor field.
(This holds also for the nonkinetic terms of ${\cal T}^{IJK}$
which we have not displayed.)
If the limit is taken in this way, the physical dimension $r$ shrinks to
zero. Presumably the interpretation of this limit
is that the squashed ${\bf CP}^2$-brane can collapse
by radiation of the antisymmetric tensor field as well as by radiation
of gravitons.
\vskip.1in
\noindent{\bf 4. Supersymmetry and the $CP^2$ solution}
\vskip .1in
The Lagrangian of the matrix theory is invariant under
supersymmetry transformations. The supersymmetry
variation of $\theta$ is given by
$$
\delta \theta \equiv K\epsilon +\rho ={{1}\over{2}} \left[{\dot X}_I\Gamma_I
+[X_I,X_J]\Gamma_{IJ}
\right]\epsilon + \rho \eqno(24)
$$
where  $\epsilon$ and $\rho$ are $16$-component spinors
of $O(9)$.

In this section, we shall examine the question of supersymmetry of the
solutions.
Within the class of
solutions considered in this paper, the potential energy, for a fixed
$R$, is finite as $s\rightarrow
\infty$ only for
(14) and (15).  Therefore we shall examine only these
cases.

Since $\delta \theta$ has $s$-dependent terms, the question of supersymmetry
is best understood by considering fermionic collective coordinates.
These are introduced by using the supersymmetry variation
(24) with the parameters $\epsilon,~\rho$ taken to be time-dependent.
Upon substitution in the Lagrangian, the term $\Tr [ \theta^T {\dot \theta}]$
generates the symplectic structure for $(\epsilon , \rho )$.
We can then construct the supersymmetry generators for fluctuations
around our solution.
If the starting configuration is supersymmetric, there will be zero modes
in the symplectic form so constructed and we will have only a smaller
number of fermionic parameters appearing in $\Tr [\theta^T {\dot \theta}]$.
Now, in the large $s$-limit, we have $\Tr [\theta^T {\dot \theta}]\sim
s^2 [ \epsilon^T K^TK {\dot \epsilon} +\rho^T {\dot \rho} ]$ which goes
to zero as $s\rightarrow \infty$ if $\delta \theta \sim s^{-1-\eta},~\eta >0$.
Thus if $\delta \theta$ vanishes faster than $1/s$, we can conclude that
the starting bosonic configuration is supersymmetric.

We now turn to the specific solutions, taking up the squashed ${\bf CP}^2$
first.
The finiteness of the kinetic energy in the large
$N$ limit requires that the leading term in $r$ must be a constant, which is
how we obtained a static solution. The equation of motion then
shows that
${\dot r}$ must go like ${1\over N}\sim {1\over s^2}$.
In other words, we can write $r= r_{0} + {{1}\over {N}} r_{1} + ...$.
The ${\dot X}_I$-term of $\delta\theta$ thus vanishes to the order
required. The vanishing of $\delta \theta$ (or the BPS-like condition)
then becomes, to leading
order,
$$
-{{8r^2_o}\over {N}}(\lambda_a L_a + \sqrt {3}\lambda_8 R_1)\epsilon +\rho=0
\eqno(25)
$$
where the set $\lambda_a,\lambda_8$, $a=1,2,3$ generate an $SU(2)\times
U(1)$ subgroup
of $SU(3)$ while the operators  $ L_a, R_1$ generate an $SU(2)_L\times U(1)_R$
subgroup of $O(4)\sim SU(2)_L\times SU(2)_R$.
In terms of $(16\times 16)~\Gamma$-matrices they are given by
$$\eqalign{
L_1&={{i}\over{4}} (\Gamma_1\Gamma_3 + \Gamma_4\Gamma_2)\cr
L_2&={{i}\over{4}} (\Gamma_1\Gamma_2 + \Gamma_3\Gamma_4)\cr
L_3&={{i}\over{4}} (\Gamma_2\Gamma_3 + \Gamma_1\Gamma_4)\cr
R_1&={{i}\over{4}} (\Gamma_1\Gamma_3 - \Gamma_4\Gamma_2)\cr}
\eqno(23)
$$
Notice that the $\epsilon$-term is of order $1/s$ since
$\lambda_a,~\lambda_8$ have eigenvalues of order $s$ and in the large $s$ limit
$N\approx \half s^2$; the $\rho$-term
is of order one. Thus condition (25) is required for supersymmetry as
explained above.
Further, in our problem
the $O(9)$ group is broken to $O(4)\times O(5)$.
With respect to this breaking, the
$16$-component spinor of $O(9)$ decomposes according to
${\underline {16}}= ((1,2),4) + ((2,1), 4)$, where $4$ denotes the spinor
of $O(5)$. In terms of the
$SU(2)_L\times U(1)_R$ subgroup generated by the $ L_a$ and $R_1$ we
have
$(1,2) = 1_1+1_{-1}$ and $(2,1)= 2_0$ where the subscripts denote the
$U(1)$ charges.
Clearly we have no singlets under $SU(2)_L\times U(1)_R$; $SU(2)_L$-singlets
necessarily carry $U(1)$ charges. Therefore the operator
$(\lambda_a L_a + \sqrt {3}\lambda_8 R_1)$ can neither annihilate $\epsilon$
nor can it be a multiple of the
unit operator in the $SU(3)$ space. Hence a nontrivial
$\epsilon$-supersymmetry cannot be compensated by a
$\rho$-transformation. Thus we have no supersymmetry.

The supersymmetry variation produces a $\theta$ of the form
$(\lambda_a L_a + \sqrt {3}\lambda_8 R_1)\epsilon$, where we set
$\rho =0$ for the moment. The contribution of this $\theta$ to the
Hamiltonian via the term $\Tr (\theta^T [X_i,\theta ])$ is zero
due to the orthogonality of the $SU(N)$ generators. In other
words, the configurations $(X_i,0)$ and $(X_i,\delta \theta )$
have the same energy. This gives a supersymmetric set of degenerate
configurations, or supermultiplets upon quantization. The starting
bosonic configuration is not supersymmetric but is part of a set of
degenerate configurations related by supersymmetry.

Consider now the solution (15). In this case also, a $\rho$-transformation
cannot compensate for an $\epsilon$-transformation and the condition for
supersymmetry becomes $f_{IJK}\Gamma_I\Gamma_J \epsilon =0$, where
$f_{IJK}$ are the structure constants of $SU(3)$. $L_K=f_{IJK}\Gamma_I\Gamma_J$
obey the ${\underline{SU(3)}}$ commutation rules
and indeed this defines an
$SU(3)$ subgroup of $O(8)$. The spinors of $O(8)$ do not contain singlets
under this $SU(3)$ and hence there is again no supersymmetry.

Although we do not have supersymmetry in the strict sense it is
perhaps interesting to note that we do have  unbroken supersymmetries
in what has been called locally flat regions in reference [6].
Again, considering the squashed ${\bf CP}^2$ for simplicity,
this can be seen by choosing
$\epsilon$ and
$\rho$ to be of the
$(1,2)$ type in
the notation introduced above. We shall choose
them to be $1_1$ of the $SU(2)\times U(1)$ group. Similar arguments will
apply to $1_{-1}$ component.
The BPS condition then  reduces to
$$
 -{{8r^2_o}\over {N}} \sqrt {3}\lambda_8 \epsilon +\rho =0 \eqno(27)
$$
To apply the notion of local flatness of [6] to our
problem we consider those regions of the diagonal matrix $\lambda_8$
where the matrix elements are close to their maximum values, which
are of the order of $s$.
In these regions $(1/s) \lambda_8$ will behave
like $1+c/s$, where $c=-{{3}\over{2}}(k+l)$ . In general
$k$ and $l$, which indicate the
number of the indices equal to $1$ and $2$
in a symmetric rank $s$ tensor
of $SU(3)$,
can range between $1$ and $s$  such that $k+l\leq s$. The
locally flat region is defined by the condition that
$k+l$ is much smaller than $s$. In such a region the BPS equation
(27) will become
$$
{{8r^2_o}\over {s}}(1+ {{c}\over {s}} )\epsilon +\rho =0 \eqno(28a)
$$
When
$c$ becomes close to $s$ we should end up in a different locally flat
region. It is clear from (28) that with an appropriate choice of
$\rho$ it is possible to compensate the supersymmetry generated by
$\epsilon$.

A more geometrical way of expressing this observation is to substitute
the semi-classical large $s$ expression for $\lambda_8$ in (27). We
then obtain the following local expression
$$
{{4r^2_o}\over {s}}~ \left[{{2 - \bx \xi}\over{1 +
\bar\xi\xi}}\right] \epsilon +\rho =0 \eqno(28b)
$$

Equation (28) corresponds to the region $\xi=0$. Other values of $\xi$ will
correspond
to different submatrices of $\lambda_8$ where it can be written as
the unit matrix plus small corrections of the order of $1/s$.

{}From this discussion we see that in the large $N$ limit we have
a kind of local supersymmetry, which cannot be extended globally. It
is of course well known that ${\bf CP}^2$ does not admit global spin
structure. In matrix theory this is a consequence of
the fact that the matrix $(1/s)\lambda_8$ cannot be globally
written as the sum
of the unit matrix and  a matrix with elements of the order of
$1/s$.
A global extension of this result can only be obtained if the
supersymmetry parameters also transform nontrivially under
the gauge group.
\vskip .1in
\noindent{\bf 5. Other Solutions}
\vskip .1in
So far we have presented a class of solutions of the form
$G/H$ for the equations of motion of M(atrix) theory.
We shall now comment on how smooth manifolds of various dimensions
smaller than 11 can emerge from Matrix theory in the large $N$ limit.

As mentioned in section 2, the condition for the solvability of the
equations of motion
for our class of anz\"atze is that $[X_J,[X_I,X_J]]$ should
belong to the same set
as the
$X_I$. The case
of ${\bf CP}^2$, squashed or otherwise,
given in section 3 can  clearly be generalized to ${\bf CP}^3$, ${\bf
CP}^4$ and
other coset spaces. For the ${\bf CP}^n$ case, the exponent $a$ in
(5) is given by
$a= {{1}\over{n}}$ and the potential energy goes like $s^{n-2} R r^4$.
Thus it is only for
$n=2$ that the potential energy becomes independent of $s$.

The required condition on the double
commutators is
satisfied if the $X_I$ span the entire Lie algebra of $G$ as well. The
equations
of motion for all these
cases are generically of the form
$$
\left({N\over R}\right)^{2a} {\ddot f} +c_2 (adj) f^3 =0 \eqno(29)
$$
where $c_2 (adj)$ represents the quadratic Casimir of $G$ in the
adjoint representation and
$r= f R^{1-a}$. Since there are only nine $X_I$, if $G$ is not a product
group, its dimension
for this type of solutions cannot exceed $9$.

First consider the case that the generators are parametrized
in terms of ${\bf CP}^n$ coordinates, as in Eqs.(11), eventhough
the $X_i$ span the whole Lie algebra of $G$.
\footnote{*}{ For all the ${\bf CP}^n$ cases the coset embdding will
produce a squashed manifold of the type we discussed in the previous section
for the case of $n=2$.}
The solutions in this case amount to different immersions of the
${\bf CP}^n$ solutions in ${\bf R}^9$.
The case of $G=SU(2)$ reproduces the spherical
membrane. In this case $a=1$, and, as noted before, both
kinetic and potential energies have
well-defined limits as $N,~R\rightarrow \infty$ with their ratio
fixed.

The semiclassical form of the solution in this case is
$$\eqalign{
Z &={X_1+iX_2 \over \sqrt{2}}
= r(t) {(s+2)\over (s+1)} { \bx\over {(1+\bx \cdot \xi )}}\approx
r(t) { \bx\over {(1+\bx \cdot \xi )}}\cr
{\zeta} &=X_3= \half r(t) {(s+2)\over (s+1)} {{(1- \bx\cdot \xi)}\over
{(1+\bx \cdot \xi )}}\approx \half r(t) ~{{(1- \bx\cdot \xi)}\over
{(1+\bx \cdot \xi )}}\cr}
\eqno(30)
$$
Clearly we have a two-sphere defined by
$$
Z\bZ +  \zeta^2 \approx {1\over 4}r(t)^2 \eqno(31)
$$
The radius of the sphere remains finite as $s\rightarrow \infty$.
Eventhough the ansatz has the full $SU(2)$-symmetry, there is a further
algebraic constraint, viz., Eq.(31), and this reduces the space of
free parameters to $SU(2)/U(1)$.

We have already considered
the case of $G=SU(3)$, which leads to the ansatz (15) obeying the algebraic
conditions (16). Also, as mentioned in section 2, the case of
$S^4$-geometry involves the coset $O(6)/O(5)$, with a further algebraic
condition which reduces the dimension to four.

There are many other interesting cases which can be considered along
these lines; for example, for $SU(2)\times SU(2)$, we can set
six of the $X_i$'s
proportional to the generators and the energies depend only on
$N/R$ as $N, R \rightarrow \infty$. This can be embedded in $U(N)$ in
a block-diagonal way by choosing the representation $(N_1,1)+(1,N_2)$
with $N=N_1+N_2$. Presumably this can give two copies of the two-brane
in some involved geometrical arrangement in ${\bf R}^9$.

The world volume geometry, it is clear for the above solutions, is tied to
the representation (11) of the generators. In order to have different
world geometries, we need to consider different
parametrizations of the group
generators, involving more world volume coordinates, for example.
Within our ansatz (5), we do not expect more possibilities.
For example, for $SU(2)$, consider the parametrization of the
generators
$$
t_i =\epsilon_{ijk} y_j {\partial \over \partial y_k}\eqno(32)
$$
which would naively suggest a three-dimensional world geometry.
However, it is consistent, within this parametrization, to set
$y_iy_i =1$, so that effectively, this reverts to the
previous case, viz., ${\bf CP}^1$ or $SU(2)/U(1)$.
Likewise, for $SU(3)$, we can consider a parametrization in terms of
eight real $y_i$'s. The quadratic and cubic invariants of $SU(3)$
allow a reduction to six world coordinates, with a local geometry
of the form ${SU(3)\over {U(1)\times U(1)}}$; in other words, we
can take $H={U(1)\times U(1)}$. It may be possible to find such a solution,
but it would be necessary to go beyond our ansatz (5).
\vskip .1in
\noindent{\bf 6. Conclusion}
\vskip .1in

 In this paper we constructed solutions to the matrix theory equations of
motion which, in the large $N$ limit, approach a continuum manifold of the
form
$G/H$. We gave simple group theoretical
description of the known spherical
solutions and showed how other
solutions of similar type can be
obtained. We then examined a particular case, namely $G= SU(3)$,
in some detail and showed how in the large $N$ limit we can endow the solution
with complex local coordinates and complex functions  which can be
expressed in terms of these coordinates.
We gave an explicit map from the envelopping algebra of $SU(3)$
to the
algebra of functions on ${\bf CP}^2$. Under this correspondence the Lie
bracket
is mapped to the Poisson
bracket on the space of functions.
This solution is not
supersymmetric, but is part of a set of degenerate configurations which are
related
by supersymmetry.
Further, in the large $N$
and finite $R$ limit in which the solution
becomes static, we argued that there is some residual supersymmetry
in small locally flat neighborhoods, where we used the
matrix theory version of local flatness along the lines of reference [6].
However, there is no
global extension of this result and hence the meaning or
significance of this symmetry is not clear.
Finally, we also showed how other solutions of similar variety or even a
slightly more general type, in which
the homogenous space $G/H$ is not necessarily symmetric, can be generated.
The equations of motion of the
M(atrix) theory seem to admit
many solutions of all
dimensionalities between one and nine.
It is still, however, an open question to find solutions to these equations
which  have a  better resemblance to the BPS-states of M-theory.

\vskip .2in
\noindent{\bf Acknowledgments:} We are grateful to the referee
for his useful criticism of an earlier
version of this paper. We also thank George Thompson for
useful discussions.  VPN was supported in part by the National Science
Foundation Grant number
PHY-9605216. He also thanks the Abdus Salam ICTP for hospitality.
\vskip .1in
\noindent{\bf References}
\vskip .1in
\item{1.} T. Banks, W. Fischler, S. Shenker and L. Susskind, hep-th 9610043;
\item{2.}L. Susskind, hep-th/9704080,  A. Sen,  hep-th/9709220,  N.
Seiberg hep-th/ 9710009;
for  recent reviews, see
T. Banks hep-th/9710231, D. Bigatti and L. Susskind hep-th/9712072,
W. Taylor IV hep-th/9801182.
\item{3.} M. Dine and A. Rajaraman, hep-th 9710174.
\item{4.} M. Douglas and H. Ooguri, hep-th 9710178.
\item{4.} B. de Wit, J. Hoppe and H. Nicolai, {\it Nucl.Phys.} {\bf B305}
(1988) 545.
\item{6.} D. Kabat and W. Taylor, hep-th 9711078;
\item{7.} S.J. Rey, hep-th 9711081.
\item{8.} T. Banks, N. Seiberg and S. Shenker, hep-th 9612157.
\item{9.} O.J. Ganor, S. Ramgoolam and W. Taylor, hep-th 9611202.
\item{10.} J. Castelino, S. Lee and W. Taylor, hep-th 9712105.
\item{11.} D. Kabat and W. Taylor, hep-th 9712185.

\end